\documentclass[12pt]{article}
\usepackage{subfloat}
\pdfoutput = 1
\begin{document}
\date{Today}
\title{{\bf{\Large  Holographic superconductors in Born-Infeld electrodynamics and external magnetic field}}}

\author{
{\bf {\normalsize Sunandan Gangopadhyay}$^{a,b,c}
$\thanks{sunandan.gangopadhyay@gmail.com}}\\
$^{a}$ {\normalsize National Institute for Theoretical Physics, Stellenbosch University, South Africa}\\
$^{b}$ {\normalsize Department of Physics, West Bengal State University, Barasat, Kolkata 700126, India}\\
$^{c}${\normalsize Visiting Associate in Inter University Centre for Astronomy \& Astrophysics,}\\
{\normalsize Pune, India}\\[0.3cm]
}
\date{}

\maketitle

\begin{abstract}
{\noindent In this paper, we analytically investigate the effect of adding an external magnetic field
in presence of Born-Infeld corrections to a holographic superconductor in the probe limit.
The technique employed is based on the matching of the solutions to the field equations near the
horizon and the asymptotic AdS region. We obtain expressions for the critical temperature and the
condensation values explicitly to all orders in the Born-Infeld parameter. 
The value of the critical magnetic field is finally obtained and is
found to get affected by the Born-Infeld parameter. }
\end{abstract}
\vskip 1cm


The formulation of the BCS theory of superconductivity over five decades ago
still remains the most successful microscopic theory describing the phenomenology
of superconductivity exhibited by a large number of metals and alloys \cite{bcs}.
However, there is a class of materials which exhibit superconductivity at high
temperatures, namely the high $T_c$ cuprates. The theoretical understanding of the
pairing mechanism leading to a charged condensate remains a mystery for these materials.
It is by now clear that these materials are strongly coupled and hence a new theoretical 
breakthrough is required to understand them.

The gauge/gravity correspondence discovered first in string theory
has proved to be an important tool to deal with strongly coupled systems 
\cite{adscft1}-\cite{adscft4}.
Recently, it has provided some excellent theoretical insights to understand the physics of high $T_c$ 
superconductors. The dual gravitational description of a superconductor observed in \cite{hs1}-\cite{hs6} involved the
mechanism of spontaneous breaking of a local $U(1)$ symmetry near the event horizon of the black hole 
resulting in the formation of a scalar hair condensate at a temperature $T$ less than a certain critical value $T_c$. 

A major aspect of superconductors is their response to magnetic field. They exhibit perfect diamagnetism
as the temperature is lowered below $T_c$ in the presence of an external magnetic field and the phenomenon
is known as Meissner effect \cite{tink}. Till date a number of numerical studies have been made to study
the effects of external magnetic field on holographic superconductors \cite{maeda}-\cite{albash} but relatively small number
of analytical invetigations have been made \cite{dib}-\cite{lala}. Further, the analyses (both in the presence
and absence of external magnetic field) are carried out mostly in the
framework of Maxwell electrodynamics \cite{hs7}-\cite{hs18}. In this paper, we present a simple analytical approach based on the matching
of the solutions to the field equations near the horizon and near the asymptotic AdS region \cite{hs8} to study the
response of a holographic superconductor to an external magnetic field in the presence of Born-Infeld corrections\footnote{The effect
of Born-Infeld corrections without an external magnetic field has already been investigated in \cite{sg1} by the Sturm-Lioville approach.}. The analysis is important as Born-Infeld electrodynamics is one of the most important non-linear
electromagnetic theory free from infinite self energies of charged point particles arising 
in the Maxwell theory. The theory is also invariant under the electromagnetic
duality transformations \cite{infeld}. Further, the investigation helps us to understand holographic superconductors 
in the framework of non-linear electrodynamics \cite{hs19}-\cite{hs22}.
The analytical expressions for the critical temperature $T_c$ and the condensation values are first obtained in the absence
of the magnetic field. The results are compared with those obtained 
from the Sturm-Liouville technique \cite{hs7},\cite{sg1}-\cite{sg5}. An important feature of
our analysis is that the calculations are performed to all orders in the Born-Infeld parameter. The magnetic field is then added
and the procedure is followed once again to obtain the value of the critical magnetic field above which the superconducting
phase dissappears. The analysis reveals that the critical magnetic field gets affected by the Born-Infeld parameter.
It is also important to note that we carry out our computations in the probe limit.

We start our construction of the holographic $s$-wave superconductor on the fixed
background of Schwarzschild-AdS spacetime.
The metric of a planar Schwarzschild-AdS black hole reads
\begin{eqnarray}
ds^2=-f(r)dt^2+\frac{1}{f(r)}dr^2+r^2(dx^2+dy^2)
\label{m1}
\end{eqnarray}
with
\begin{eqnarray}
f(r)=r^{2}-\frac{r_{+}^3}{r}
\label{metric}
\end{eqnarray}
in units in which the AdS radius is unity, i.e. $l=1$.
The Hawking temperature is related to the horizon radius ($r_+$) as
\begin{eqnarray}
T=\frac{3r_+}{4\pi}~.
\label{temp}
\end{eqnarray}
We now consider a charged complex scalar field minimally coupled to the Maxwell field $A_{\mu}$ in this fixed background. 
The corresponding Lagrangian density can be expressed as
\begin{eqnarray}
\mathcal{L}=\mathcal{L}_{BI}-
|\nabla_{\mu}\psi-iqA_{\mu}\psi|^2-m^2|\psi|^2
\label{m10}
\end{eqnarray}
where $\psi$ is a charged complex scalar field, $\mathcal{L}_{BI}$
is the Lagrangian density of the Born-Infeld electrodynamics
\begin{eqnarray}
\mathcal{L}_{BI}=\frac{1}{b}\bigg(1-\sqrt{1+\frac{b F}{2}}\bigg)
\label{m11}
\end{eqnarray}
where $F\equiv F_{\mu\nu}F^{\mu\nu}$ and $F_{\mu\nu}$ is the
BI electromagnetic tensor and $b$ is the BI coupling parameter. 

\noindent In order to solve the equations of motion for the electromagnetic field and the complex scalar field,
we make the following ansatz \cite{hs6}
\begin{eqnarray}
A_{\mu}=(\phi(r),0,0,0),\;\;\;\;\psi=\psi(r)
\label{vector}
\end{eqnarray}
which finally yields the following set of equations of motion for the electrical scalar potential $\phi(r)$
and the complex scalar field $\psi(r)$:
\begin{eqnarray}
\phi''(r)+\frac{2}{r}\phi'(r)\bigg(1-b\phi'^2 (r)\bigg)
-\frac{2\psi^2 (r)}{f}\phi(r)\bigg(1-b\phi'^2 (r)\bigg)^{3/2}=0\label{e2}
\end{eqnarray}
and 
\begin{eqnarray}
\psi^{''}(r)+\left(\frac{f'}{f}+\frac{2}{r}\right)\psi'(r)
+\left(\frac{\phi^{2}(r)}{f^2}-\frac{m^2}{f}\right)\psi(r)=0\label{e1}
\end{eqnarray}
where prime denotes derivative with respect to $r$. In order to solve the
non-linear equations (\ref{e2}) and (\ref{e1}), we need
to seek the boundary condition for $\phi$ and $\psi$ near the black
hole horizon $r\sim r_+$ and at the spatial infinite
$r\rightarrow\infty$. The regularity condition at the horizon gives
the boundary conditions $\phi(r_+)=0$ and $\psi'(r_+)=-\frac{2}{3r_{+}}\psi(r_+)$. 
The second relation follows from eq.(\ref{e1}) using the fact that
$f(r_+)=0=\phi(r_+)$ and $f'(r_+)=3r_{+}$.

\noindent Under the change of coordinates $z=r_{+}/r$ and setting $m^2=-2$,  the field equations become
\begin{eqnarray}
\phi''(z)+\frac{2bz^3}{r_{+}^2 }\phi'^{3}(z)-\frac{2r_{+}^2 \psi^2 (z)}{z^4 f(z)}
\left(1-\frac{bz^4}{r_{+}^2}\phi'^{2}(z)\right)^{3/2}\phi(z) =0\label{e1aa}
\end{eqnarray}
\begin{eqnarray}
\psi''(z)+\frac{f'(z)}{f(z)}\psi'(z)+\frac{2r_{+}^2}{z^4 f(z)}\psi(z)
+\frac{r_{+}^2 \phi^{2}(z)}{z^4 f^{2}(z)}\psi(z)=0
\label{e1a}
\end{eqnarray}
where prime now denotes derivative with respect to $z$. These equations are to be solved in the
interval $(0, 1)$, where $z=1$ is the horizon and $z=0$ is the boundary.
The boundary conditions for $\phi$ and $\psi$ in terms of $z$ coordinates now become $\phi(z=1)=0$ and  
$\psi'(1)=\frac{2}{3}\psi(1)$. The second relation follows from eq.(\ref{e1a}) using the fact that
$f(1)=0=\phi(1)$ and $f'(1)=-3r_{+}^2$.

\noindent  The asymptotic boundary conditions for
the scalar potential $\phi(z)$ and the scalar field $\psi(z)$ turn out to be
\begin{eqnarray}
\phi(z)&\approx&\mu-\frac{\rho}{r_{+}}z\nonumber\\
\psi(z)&\approx&J_{-}z+J_{+}z^2~.
\label{b2}
\end{eqnarray}
The coefficients $\mu$ and $\rho$ are 
interpreted as the chemical potential and the charge density of the dual theory
on the boundary. In the subsequent analysis, we shall set $J_{+}=0$\footnote{This is the condition with which the
holographic superconductor has been analysed in \cite{sg1} using the Sturm-Liouville technique.}. 
The condensation operator $\mathcal{O}$ is related to $J_{-}$ as $\mathcal{O}=\sqrt{2}r_{+}J_{-}$.

\noindent With the above relations in place, we now proceed to present a simple analytic approach
to obtain expressions for the critical temperature and values for the condensation.
The method involves finding an approximate solution around $z=1$ and $z=0$ using
Taylor expansion and then matching these solutions between $z=1$ and $z=0$ \cite{hs8}.

\noindent Near the horizon $z=1$, we expand $\phi(z)$ and $\psi(z)$ as 
\begin{eqnarray}
\label{ma100}
\phi(z)&=&\phi(1)-\phi'(1)(1-z)+\frac{1}{2}\phi''(1)(1-z)^2 +...\nonumber\\
&\approx&-\phi'(1)(1-z)+\frac{1}{2}\phi''(1)(1-z)^2\\
\psi(z)&=&\psi(1)-\psi'(1)(1-z)+\frac{1}{2}\psi''(1)(1-z)^2 +...
\label{ma8}
\end{eqnarray}
where we have used the boundary condition $\phi(1)=0$ in the second line of eq.(\ref{ma100}).
We also set $\phi'(1)<0$ and $\psi(1)>0$ without loss of generality.

\noindent Now from eqs(\ref{e1aa}, \ref{e1a}) and using the regularity conditions for $\phi$ and $\psi$
and the relations $f'(1)=-3r_{+}^2$ and $f''(1)=6r_{+}^2$, we compute the second derivatives of 
$\phi(z)$ and $\psi(z)$ exactly at the horizon $z=1$\footnote{Eq.(\ref{za2}) differs from the corresponding equation in \cite{dib}.}:
\begin{eqnarray}
\label{za1}
\phi''(1)&=&-\frac{2b}{r_{+}^2}\phi'^{3}(1)-\frac{2}{3}\psi^{2}(1)\phi'(1)\left(1-\frac{b}{r_{+}^2}\phi'^{2}(1)\right)^{3/2}\\
\psi''(1)&=&-\frac{4}{9}\psi(1)-\frac{\phi'^{2}(1)\psi(1)}{18r_{+}^2}~.
\label{za2}
\end{eqnarray}
Substituting eq.(\ref{za1}) in eq.(\ref{ma100}) and eq.(\ref{za2}) in eq.(\ref{ma8}), we get
\begin{eqnarray}
\label{ma9}
\phi(z)&\approx&-\phi'(1)(1-z)-\left[\frac{b}{r_{+}^2}\phi'^{2}(1)+\frac{1}{3}\psi^{2}(1)\left(1-\frac{b}{r_{+}^2}\phi'^{2}(1)\right)^{3/2}\right]\phi'(1)(1-z)^2\\
\psi(z)&\approx&\psi(1)-\psi'(1)(1-z)-
\left[\frac{2}{9}\psi(1)+\frac{\phi'^{2}(1)\psi(1)}{36 r_{+}^2}\right](1-z)^2 \nonumber\\
&=&\frac{1}{3}\psi(1)+\frac{2}{3}\psi(1)z-\left[\frac{2}{9}\psi(1)+\frac{\phi'^{2}(1)\psi(1)}{36 r_{+}^2}\right](1-z)^2 
\label{ma10}
\end{eqnarray}
where in the second line of eq.(\ref{ma10}), we have used the regularity condition for $\psi$ at the horizon $z=1$.

\noindent Now we match the above solutions with eq.(\ref{b2}) (with $J_{+}=0$)
 at some intermediate point $z=z_m$. This leads to the following equations:
\begin{eqnarray}
\label{co1}
\mu-\frac{\rho}{r_{+}}z_m&=&\beta(1-z_m)+\beta\left[\frac{b\beta^2}{r_{+}^2}+\frac{\alpha^2}{3}\left(1-\frac{b\beta^{2}}{r_{+}^2}\right)^{3/2}\right](1-z_m)^2\\
\label{co1a}
\frac{\rho}{r_{+}}&=&\beta+2\beta\left[\frac{b\beta^2}{r_{+}^2}+\frac{\alpha^2}{3}\left(1-\frac{b\beta^{2}}{r_{+}^2}\right)^{3/2}\right](1-z_m)\\
\label{co2}
J_{-}z_m&=&\frac{\alpha}{3}+\frac{2\alpha}{3}z_m-\frac{\alpha}{9}\left[2+\frac{\beta^{2}}{4 r_{+}^2}\right]
(1-z_m)^2\\
\label{co2a}
J_{-}&=&\frac{2\alpha}{3}+\frac{\alpha}{9}\left[4+\frac{\beta^{2}}{2 r_{+}^2}\right](1-z_m)
\end{eqnarray}
where we have set $\beta=-\phi'(1)$ and $\alpha=\psi(1)$.

\noindent From eq.(\ref{co1a}) and using eq.(\ref{temp}), we obtain
\begin{eqnarray}
\alpha^2 =\frac{3[1+2b\tilde{\beta}^2 (1-z_m)]}{2(1-z_m)(1-b\tilde{\beta}^2)^{3/2}}\left(\frac{T_{c}^2}{T^2}-1\right)
\label{tc1}
\end{eqnarray}
where $\tilde{\beta}=\frac{\beta}{r_+}$ and $T_c$ is given by
\begin{eqnarray}
T_c =\kappa\sqrt{\rho}
\label{tc2}
\end{eqnarray}
where
\begin{eqnarray}
\kappa =\frac{3}{4\pi}\frac{1}{\sqrt{\tilde\beta [1+2b\tilde{\beta}^2 (1-z_m)]}}~.
\label{tc2nm}
\end{eqnarray}
For $T\sim T_c$, eq.(\ref{tc1}) leads to
\begin{eqnarray}
\alpha=\sqrt{\frac{3[1+2b\tilde{\beta}^2 (1-z_m)]}{(1-z_m)(1-b\tilde{\beta}^2)^{3/2}}}
\sqrt{1-\frac{T}{T_c}}~.
\label{tc3}
\end{eqnarray}
It is to be noted that the expressions for $\alpha$ and $T_c$ are exact to all orders in the Born-Infeld parameter $b$.

\noindent From eq(s)(\ref{co2}, \ref{co2a}), we obtain
\begin{eqnarray}
\tilde\beta=2\sqrt{\frac{1+2z_{m}^2}{1-z_{m}^2}}~~;~~
J_{-}=\frac{2\alpha(2+z_m)}{3(1+z_m)}~.
\label{tc4}
\end{eqnarray}
The condensation operator may now be computed from eqs(\ref{tc3}, \ref{tc4}):
\begin{eqnarray}
\mathcal{O}=\sqrt{2}r_{+}J_{-}=\gamma T_{c}\sqrt{1-\frac{T}{T_c}}
\label{tc5}
\end{eqnarray}
where $\gamma$ is given by
\begin{eqnarray}
\gamma=\frac{8\sqrt{2}\pi (2+z_m)}{9(1+z_m)}
\sqrt{\frac{3[1+2b\tilde{\beta}^2 (1-z_m)]}{(1-z_m)(1-b\tilde{\beta}^2)^{3/2}}}~.
\label{tc6}
\end{eqnarray}
We now summarize the results obtained above in a table.
\begin{table}[htb]
\caption{A comparison of the analytical [matching (with $z_m =0.1$) and Sturm-Liouville (SL) methods] and numerical results for the critical temperature }   
\centering                          
\begin{tabular}{c c c c c c c}            
\hline\hline                        
$b$ & $\kappa_{matching}$  & $\kappa_{SL}$ &$\kappa_{numerical}$ &  \\ [0.05ex]
\hline
0 & 0.168 & 0.225 & 0.226 \\
0.1 & 0.127 & 0.223 & 0.224 \\                              
[0.05ex]  
\hline                              
\end{tabular}\label{E1}  
\end{table}
The comparison with numerical results show that the Sturm-Liouville method yields better results than the matching method.
This feature was also found to be true in \cite{sg3} where a comparison of the two techniques had been made
with the numerical results.

\noindent We now move on to investigate the effect of an external static magnetic field in the model of
the holographic superconductor considered so far. This is done by adding a magnetic field in the bulk.
The asymptotic value of this magnetic field corresponds to a magnetic field added to the boundary field theory
by the gauge gravity correspondence. To proceed further, we make the following ansatz \cite{albash}
\begin{eqnarray}
A_t = \phi(z)~, ~A_y =Bx~,~\psi=\psi(x, z).
\label{mg1}
\end{eqnarray}
This leads to the following equation for the scalar field $\psi$
\begin{eqnarray}
\psi''(x, z)+\frac{f'(z)}{f(z)}\psi'(x, z)+\frac{2r_{+}^2 \psi(x, z)}{z^4 f(z)}
+\frac{r_{+}^2 \phi^{2}(z)\psi(x, z)}{z^4 f^{2}(z)}+\frac{1}{z^2 f(z)}(\partial^{2}_{x}\psi -B^2 x^2 \psi)=0.
\label{mg2}
\end{eqnarray}
To solve this equation, we take the following separable form for $\psi$
\begin{eqnarray}
\psi(x, z)=X(x)R(z).
\label{mg3}
\end{eqnarray}
Substituting this in eq.(\ref{mg2}) leads to
\begin{eqnarray}
z^2 f(z)\left[\frac{R''}{R}+\frac{f'}{f}\frac{R'}{R}+\frac{r_{+}^2 \phi^2}{z^4 f^2}+\frac{2r_{+}^2}{z^4 f}\right]-\left[-\frac{X''}{X}+B^2 x^2\right]=0.
\label{mg4}
\end{eqnarray}
The equation for $X(x)$ is readily identified as the Schr\"{o}dinger equation in one dimension with 
a $B$ dependent frequency \cite{albash}
\begin{eqnarray}
-X''(x)+B^2 x^2 X(x)=\lambda_n BX(x)
\label{mg5}
\end{eqnarray}
where $\lambda_n =2n+1$ is a constant. Considering the lowest mode ($n=0$) expected to be the most stable \cite{albash}
leads to the following equation for $R(z)$:
\begin{eqnarray}
R''(z)+\frac{f'(z)}{f(z)}R'(z)+\frac{2r_{+}^2 R(z)}{z^4 f(z)}+\frac{r_{+}^2 \phi^2 (z) R(z)}{z^4 f^2 (z)}
=\frac{B R(z)}{z^2 f(z)}~.
\label{mg6}
\end{eqnarray}
From this equation, using the regularity condition for $\phi$ and $f'(1)=-3r_{+}^2$, 
one can obtain the following relation at the horizon ($z=1$):
\begin{eqnarray}
R'(1)=\left(\frac{2}{3}-\frac{B}{3r_{+}^2}\right)R(1)~.
\label{mg7}
\end{eqnarray}
The asymptotic solution for eq.(\ref{mg6}) can be written as
\begin{eqnarray}
R(z)=J_{-}z+J_{+}z^2
\label{mg8}
\end{eqnarray}
where according to our previous choice $J_{+}=0$.

\noindent The aim is to find the value of the (critical) magnetic field for which the condensate vanishes.
To proceed, once again we take recourse to the matching method \cite{hs8}.

\noindent Near the horizon $z=1$, we expand $R(z)$  
\begin{eqnarray}
\label{ma1}
R(z)=R(1)-R'(1)(1-z)+\frac{1}{2}R''(1)(1-z)^2 +...~.
\label{mg9}
\end{eqnarray}
From eq.(\ref{mg6}) and using the regularity condition for $\phi$, $f'(1)=-3r_{+}^2$, 
$f''(1)=6r_{+}^2$ and eq.(\ref{mg7}), we obtain\footnote{Eq.(\ref{mg10}) differs from the corresponding equation in \cite{dib}.}
\begin{eqnarray}
\label{ma1}
R''(1)=\left[-\frac{4}{9}-\frac{2B}{9r_{+}^2}+\frac{B^2}{18r_{+}^4}-\frac{\phi'^2(1)}{18r_{+}^2}\right]R(1)~.
\label{mg10}
\end{eqnarray}
Substituting this in eq.(\ref{mg9}) and using eq.(\ref{mg7}), we get
\begin{eqnarray}
R(z)\approx\frac{1}{3}R(1)+\frac{2}{3}R(1)z+\frac{BR(1)}{3r_{+}^2}(1-z)+\frac{1}{2}\left[-\frac{4}{9}-\frac{2B}{9r_{+}^2}+\frac{B^2}{18r_{+}^4}-\frac{\phi'^2(1)}{18r_{+}^2}\right]R(1)(1-z)^2 ~.
\label{mg11}
\end{eqnarray}
Matching this solution with eq.(\ref{mg8}) (with $J_{+}=0$) at some intermediate point $z=z_m$ leads to
the following equations:
\begin{eqnarray}
\label{mg12}
J_{-}z_{m}&=&\frac{1}{3}R(1)+\frac{2}{3}R(1)z_m+\frac{BR(1)}{3r_{+}^2}(1-z_m)\nonumber\\
&&+\frac{1}{2}\left[-\frac{4}{9}-\frac{2B}{9r_{+}^2}+\frac{B^2}{18r_{+}^4}-\frac{\phi'^2(1)}{18r_{+}^2}\right]R(1)(1-z_m)^2 \\
J_{-}&=&\frac{2}{3}R(1)-\frac{BR(1)}{3r_{+}^2}z_m+\left[\frac{4}{9}+\frac{2B}{9r_{+}^2}-\frac{B^2}{18r_{+}^4}+\frac{\phi'^2(1)}{18r_{+}^2}\right]R(1)(1-z_m)~.
\label{mg13}
\end{eqnarray}
The above set of equations yield the following quadratic equation in $B$:
\begin{eqnarray}
B^2 +4r_{+}^2\left(\frac{2+z_{m}^2}{1-z_{m}^2}\right)B+4r_{+}^4\left(\frac{1+2z_{m}^2}{1-z_{m}^2}\right)
-\phi'^2(1)r_{+}^2=0
\label{mg14}
\end{eqnarray}
solving which we get
\begin{eqnarray}
B=\sqrt{\phi'^2(1)r_{+}^2 +12r_{+}^4 \frac{(1+z_{m}^2 +z_{m}^4)}{(1-z_{m}^2)^2}}-2r_{+}^2\left(\frac{2+z_{m}^2}{1-z_{m}^2}\right)~.
\label{mg15}
\end{eqnarray}
Now we consider the situation where the value of the magnetic field is very close to the
critical value which in turn implies a vanishingly small condensate. Hence one can neglect all the quadratic
terms in $\psi$. With this physical argument, eq.(\ref{e1aa}) simplifies to
\begin{eqnarray}
\phi''(z)+\frac{2bz^3}{r_{+}^2 }\phi'^{3}(z)=0.
\label{mg16}
\end{eqnarray}
Integrating this equation in the interval $[1, z]$ and using the asymptotic boundary condition for $\phi$ eq.(\ref{b2})
leads to \cite{sg1}
\begin{eqnarray}
\phi'(z)=-\frac{\lambda r_{+}}{\sqrt{1+b\lambda^2 z^4}}~;~\lambda=\frac{\rho}{r_{+}^2}.
\label{mg17}
\end{eqnarray}
Evaluating $\phi'(z)$ at $z=1$ and squarring gives
\begin{eqnarray}
\phi'^2(1)r_{+}^2 =\frac{\rho^2}{1+b\frac{\rho^2}{r_{4}^4}}~.
\label{mg18}
\end{eqnarray}
Now from eq(s)(\ref{tc2}, \ref{tc2nm}) (using the value of $\tilde\beta$ from eq.(\ref{tc4})), we have
\begin{eqnarray}
\rho=\frac{16\pi^2}{9}\tilde{\beta}\left[1+8b\frac{(1+2z_{m}^2)}{(1+z_m)}\right]T_{c}^{2}(0)
\label{mg19}
\end{eqnarray}
where $T_{c}(0)$ is the critical temperature at zero magnetic field.

\noindent Eq(s)(\ref{mg15}, \ref{mg18}, \ref{mg19}) yields the value of the critical magnetic field to be
\begin{eqnarray}
B_c&=&\frac{16\pi^2}{9}\tilde{\beta}\left[1+8b\frac{(1+2z_{m}^2)}{(1+z_m)}\right]T_{c}^{2}(0)\nonumber\\
&&\times\left[\sqrt{\frac{1}{1+b\tilde{\beta}^2 f^{2}(b, z_m)\left(\frac{T_{c}(0)}{T}\right)^2}+\frac{A_1}{\tilde{\beta}^2}\left(\frac{T}{T_{c}(0)}\right)^4}
-\frac{A_2}{\tilde\beta}\left(\frac{T}{T_{c}(0)}\right)^2\right]
\label{mg20}
\end{eqnarray}
where
\begin{eqnarray}
f(b, z_m)&=&1+8b\frac{(1+2z_{m}^2)}{(1+z_m)}\nonumber\\
A_1&=&\frac{12(1+z_{m}^2 +z_{m}^4)}{f^2(b, z_m)(1-z_{m}^2)^2 }\nonumber\\
A_2 &=&\frac{2}{f(b, z_m)}\left(\frac{2+z_{m}^2}{1-z_{m}^2}\right). 
\label{mg21}
\end{eqnarray}
The above result clearly reveals the dependence of the critical magnetic field on the Born-Infeld parameter $b$.
The result is also exact in all orders in $b$.

\noindent We conclude by summarizing the results that we have obtained.
We applied the matching method to analytically study the effect of adding an external magnetic field
in presence of Born-Infeld corrections to a holographic superconductor. The calculations are carried out in the probe limit.
Expressions for the critical temperature and the
condensation values are obtained explicitly to all orders in the Born-Infeld parameter. The results are compared with
those obtained from the Sturm-Liouville technique. The comparison of the analytical results from the two approaches
with the numerical results clearly indicate that the Sturm-Liouville technique yields better results. 
The value of the critical magnetic field is finally obtained using the matching technique and is
found to get affected by the Born-Infeld parameter.







\end{document}